\documentstyle[preprint,aps]{revtex}

\begin{document}
\thispagestyle{empty}
\begin{titlepage}
\rightline{CAS-HEP-T-95-08/002}
\rightline{DOE-ER/01545-655}
\vspace{3cm}
\begin{center}
{\LARGE\bf  Low Energy Phenomena  \\
in a Model With Symmetry Group \\ SUSY
$SO(10)\times \Delta (48) \times U(1)$}
\end{center}
\bigskip
\begin{center}
{\large\bf K.C. Chou$^{1}$ and Y.L. Wu$^{2}$} \\
$^{1}$Chinese Academy of Sciences, Beijing 100864,  China
 \\ $^{2}$Department of Physics, \ Ohio State  University \\ Columbus,
 Ohio 43210,\ U.S.A.
\end{center}
\vspace{4cm}
\begin{center}
 To be published in {\large\bf Scientia Sinica}, 1995; hep-ph/9508402
\end{center}

\end{titlepage}
\draft
\preprint{CAS-HEP-T-95-08/002}
\title{Low Energy Phenomena in a Model With Symmetry Group SUSY
$SO(10)\times \Delta (48) \times U(1)$}
\author{K.C. Chou$^{1}$ and Y.L. Wu$^{2}$\footnote{
supported in part by Department of Energy Grant\# DOE/ER/01545-655}}
\address{ $^{1}$Chinese Academy of Sciences, Beijing 100864,  China
 \\ $^{2}$Department of Physics, \ Ohio State  University \\ Columbus,
 Ohio 43210,\ U.S.A. }
\date{August 1995, hep-ph/9508402}
\maketitle

\begin{abstract}
  Fermion masses and mixing angles including that of neutrinos are studied
in a model with symmetry group SUSY $SO(10)\times \Delta (48)
\times U(1)$. Universality of Yukawa coupling
of superfields is assumed. The resulting texture of mass matrices in the low
energy region depends only on a single coupling constant and VEVs
caused by necessary symmetry breaking. 13 parameters involving masses and
mixing angles in the quark and charged lepton sector are successfully
described by only five parameters with two of them determined by the scales
of U(1), SO(10) and SU(5) symmetry breaking compatible with the requirement
of grand unification and proton decay.
The neutrino masses and mixing angles in the leptonic sector are also
determined with the addition of a Majorana coupling term.
It is found that LSND
$\bar{\nu}_{\mu}\rightarrow \bar{\nu}_{e}$ events,
atmospheric neutrino deficit and the mass limit put by hot dark matter can
be naturally explained. Solar neutrino puzzle can be solved only by introducing
sterile neutrino with one additional parameter.
More precise measurements of $\alpha_{s}(M_{Z})$,
$V_{cb}$, $V_{ub}/V_{cb}$, $m_{b}$, $m_{t}$, as well as various
CP violation and neutrino oscillation experiments will provide
crucial tests of the present model.
\end{abstract}

\newpage

\narrowtext


\section{Introduction}

The standard model (SM) is a great success. To understand the origin
of the 18 free parameters(or 25 if neutrinos are massive) is a big challenge
to high energy physics. Many efforts have been made along this direction. It
was first observed by Gatto {\it et al}, Cabbibo and Maiani\cite{CM} that the
Cabbibo angle is close to $\sqrt{
m_{d}/m_{s}}$. This observation initiated the investigation of the
texture structure with zero elements \cite{ZERO}
in the fermion Yukawa coupling matrices. A general analysis
and review of the previous studies on the texture structure was given by Raby
in \cite{RABY}. In \cite{OPERATOR,HR} Anderson {\it et al.}
presented an interesting model based on SUSY
$SO(10)$ and $U(1)$ family symmetries
with two zero textures  '11' and '13' followed naturally. Though the texture
'22' and '32' are not unique they could fit successfully the 13 observables
in the quark and charged lepton sector with only six
parameters\cite{OPERATOR}.

   In this paper we will follow their general considerations and make the
following modifications:

 1) We will use a discrete dihedral group $\Delta (3n^{2})$ with $n=4$,
a subgroup of SU(3),  as  our family group instead of U(1) used in \cite{HR}.
This kind of dihedral group was first used by Kaplan and Schmaltz \cite{KS}
with $n=5$. This group has only triplet and
singlet irreducible representations, which is well suited for our purposes.

 2) We will assume universality of Yukawa coupling before symmetry breaking
so as to reduce the possible free parameters. In this kind of theories there
are very rich structures above the GUT scale with many heavy fermions and
scalars. All heavy fields must have some reasons to exist and interact which
we do not understand at this moment. So we will just take the universality of
coupling constants at the GUT scale as a working assumption and not
worry about the possible
radiative effects. If the phenomenology is all right, one has to be more
serious to find a deeper reason for it.

 3) We shall Choose some symmetry breaking directions different
from\cite{OPERATOR,HR} to ensure
the needed Clebsch coefficients in order to eliminate further
arbitrariness of the parameters.

 Our paper is orginazied as follows: In section 2, we will present the
results of the Yukawa coupling matrices. The resulting masses and CKM
quark mixings are also presented. In section 3, neutrino masses and
CKM-type mixings in the lepton sector are presented. All existing
neutrino experiments are discussed and found to be understandable in the
present model.  In section 4, the model with superfields and
superpotential is explicitly presented. Conclusions and remarks
are presented in the last section.

\section{Yukawa Coupling Matrices}

 With the above considerations, a model based on group SUSY $SO(10)\times
\Delta(48) \times U(1)$ with a single coupling constant is
constructed. Here $U(1)$ is family-independent and  introduced to distinguish
various fields which belong to the same representations of
$SO(10)\times \Delta (48)$. Yukawa coupling matrices which determine the masses
and mixings
of all quarks and charged leptons are obtained by carefully choosing the
structure of the physical vacuum. We find
\begin{equation}
\Gamma_{f}^{G} = \frac{2}{3}\lambda_{H}
\left( \begin{array}{ccc}
0  &  \frac{3}{2}z_{f} \epsilon_{P}^{2} &   0   \\
\frac{3}{2}z_{f} \epsilon_{P}^{2} &  3 y_{f}
\epsilon_{G}^{2} e^{i\phi}
& \frac{\sqrt{3}}{2}x_{f}\epsilon_{G}^{2}  \\
0  &  \frac{\sqrt{3}}{2}x_{f}\epsilon_{G}^{2}  &  w_{f}
\end{array} \right)
\end{equation}
for $f=u,d,e$, and
\begin{equation}
\Gamma_{\nu}^{G} = \frac{2}{3}\lambda_{H}
\left( \begin{array}{ccc}
0  &  \frac{3}{2}\frac{1}{|z_{\nu}|} &   0   \\
\frac{3}{2}\frac{1}{|z_{\nu}|}  &  3 \frac{y_{\nu}}{z_{\nu}^{2}}
\frac{\epsilon_{G}^{2}}{\epsilon_{P}^{2}} e^{i\phi}
& \frac{\sqrt{3}}{2}\frac{x_{\nu}}{|z_{\nu}|}
\frac{\epsilon_{G}^{2}}{\epsilon_{P}}  \\
0  &  \frac{\sqrt{3}}{2}\frac{x_{\nu}}{|z_{\nu}|}
\frac{\epsilon_{G}^{2}}{\epsilon_{P}}  &  w_{f}
\end{array} \right)
\end{equation}
for Dirac type neutrino coupling.
$\lambda_{H}$ is the universal coupling constant expected to be
of order one.  $\epsilon_{G}\equiv v_{5}/v_{10}$ and
$\epsilon_{P}\equiv v_{5}/\bar{M}_{P}$  with $\bar{M}_{P}$, $v_{10}$ and
$v_{5}$ being the VEVs for U(1), SO(10) and SU(5) symmetry breaking
respectively. $x_{f}$, $y_{f}$, $z_{f}$ and $w_{f}$ $(f = u, d, e, \nu)$
are the Clebsch factors of $SO(10)$ determined by the
directions of symmetry breaking of the adjoints 45s.
The following three directions have been chosen for symmetry breaking, namely:
$<A_{X}>=v_{10}\  diag. (2,\ 2,\ 2,\ 2,\ 2)\otimes \tau_{2}$;
$<A_{z}> =v_{5}\  diag. (\frac{2}{3},\ \frac{2}{3},\ \frac{2}{3},\
-2,\ -2)\otimes \tau_{2}$ and  $<A_{u}>=v_{5}\
diag. (\frac{2}{3},\ \frac{2}{3},\ \frac{2}{3},\
 \frac{1}{3},\  \frac{1}{3})\otimes \tau_{2}$.
The resulting Clebsch  factors are:
$w_{u}=w_{d}=w_{e}=w_{\nu} =1$;
$x_{u}= -7/9$, $x_{d}= -5/27$, $x_{e}=1$, $x_{\nu} = -1/15$;
$y_{u}=0$, $y_{d}=y_{e}/3=2/27$, $y_{\nu} = 4/45$; $z_{u}=1$,
$z_{d}=z_{e}= -27$, $z_{\nu} = -15^3 = -3375$.
$\phi$ is the physical CP phase\footnote{ We have rotated
away other possible phases by a phase redefinition of the fermion fields.}
arising from the VEVs.
The Clebsch factors associated with the symmetry breaking directions can be
easily read off from effective operators which are obtained when the heavy
fermion pairs are integrated out and decoupled
\begin{eqnarray}
W_{33} & = & (\frac{2}{3}\lambda_{H})\  \frac{1}{2}
\ 16_{3} \ 10_{1}\  16_{3}  \nonumber \\
W_{32} & = & (\frac{2}{3}\lambda_{H}) \frac{\sqrt{3}}{2}
\epsilon_{G}^{2}\ 16_{3} (\frac{A_{z}}{v_{5}})(\frac{v_{10}}{A_{X}})\ 10_{1}\
(\frac{v_{10}}{A_{X}}) (\frac{A_{z}}{v_{5}}) \frac{1}{\sqrt{1 +
\epsilon_{P}^{2}(\frac{A_{X}}{v_{10}})^{6}}} 16_{2}  \nonumber \\
W_{22} & = & (\frac{2}{3}\lambda_{H}) \frac{3}{2} \epsilon_{G}^{2}
\ 16_{2}\frac{1}{\sqrt{1 + \epsilon_{P}^{2}(\frac{A_{X}}{v_{10}})^{6}}}
 (\frac{A_{u}}{v_{5}})(\frac{v_{10}}{A_{X}})\ 10_{1}\
(\frac{v_{10}}{A_{X}}) (\frac{A_{u}}{v_{5}})
\frac{1}{\sqrt{1 + \epsilon_{P}^{2}(\frac{A_{X}}{v_{10}})^{6}}}16_{2}  \\
W_{12} & = & (\frac{2}{3}\lambda_{H}) \frac{3}{2}  \epsilon_{P}^{2} \
16_{1} \frac{1}{\sqrt{1 + \epsilon_{P}^{2}(\frac{A_{X}}{v_{10}})^{6}}}
(\frac{A_{X}}{v_{10}})^{3}\ 10_{1}\ (\frac{A_{X}}{v_{10}})^{3}
\frac{1}{\sqrt{1 + \epsilon_{P}^{2}(\frac{A_{X}}{v_{10}})^{6}}} 16_{2}
\nonumber
\end{eqnarray}
where the factor $1/\sqrt{1 + \epsilon_{P}^{2}(\frac{A_{X}}{v_{10}})^{6}}$
arises from mixing.  The $\epsilon_{P}^{2}$ term in the
square root is negligible for quarks and charged leptons,
but it becomes dominant for the neutrinos
due to the large Clebsch factor $z_{\nu}$. In obtaining the
$\Gamma_{f}^{G}$ matrices,
some small terms  arising from mixings between the chiral
fermion $16_{i}$ and  heavy fermion pairs $\psi_{j} (\bar{\psi}_{j})$ are
neglected. They are  expected to change the numerical results no more than
a few percent. The factor $1/\sqrt{3}$ associated with the third
family is due to the maximum mixing between the third family fermion and
heavy fermions.  This set of effective operators which lead to
the above given Yukawa coupling matrices $\Gamma_{f}^{G}$ is quite unique.
Uniqueness of the structure of operator $W_{12}$
was first observed by Anderson {\it et al} \cite{OPERATOR}
from the mass ratios of
$m_{e}/m_{\mu}$ and $m_{d}/m_{s}$.
The effective operator $W_{33}$ is also fixed at the GUT
scale\cite{THIRD,THRESHOLD,OPERATOR} for the case of large $\tan \beta$.
There is only one candidate of effective operator $W_{22}$,
when the direction of breaking is chosen to be $A_{u}$,
with  Clebsch factors satisfying
$y_{u}:y_{d}:y_{e} = 0:1:3$\cite{GJ} so as to obtain a
correct mass ratio $m_{\mu}/m_{s}$.  The three parameters $\lambda_{H}$,
$\epsilon_{G}$ and $\epsilon_{P}$
are determined by the three measured mass ratios $m_{b}/m_{\tau}$,
$m_{\mu}/m_{\tau}$ and $m_{e}/m_{\tau}$. Thus,
the mass ratio $m_{c}/m_{t}$ and the CKM mixing elements
$V_{cb}$ and $V_{ub}$ put strong constriant to an unique choice of the
symmetry breaking direction $A_{z}$ for effective operator $W_{32}$.
Unlike many other models in which  $W_{33}$ is assumed to be
a renormalizable interaction before symmetry breaking, the Yukawa couplings
of all the quarks and leptons (both heavy and light) in the present model
are generated at the GUT scale after the breakdown of the family group and
SO(10). Therefore, initial
conditions of renormalization group (RG) evolution  will be set at the
GUT scale for all the quark and lepton Yukawa
couplings. Consequently, one could avoid the possible Landau pole and flavor
changing problems encountered in many other models due to RG running
of the third family Yukawa couplings from the GUT scale to
the Planck scale. The hierarchy among the three families is described
by the two ratios $\epsilon_{G}$ and $\epsilon_{P}$.
Mass splittings between  quarks and leptons as well as between the
up and down quarks are determined by the Clebsch factors of SO(10).
{}From the GUT scale down to low energies, Renormalization Group (RG)
evolution have been taken into account.
Top-bottom splitting in the present model is mainly
attributed to the hierarchy of the VEVs $v_{1}$ and $v_{2}$ of
the two light Higgs doublets in the weak scale.

 An adjoint 45 $A_{X}$ and a 16-D representation
Higgs field $\Phi$ ($\bar{\Phi}$)
are needed for breaking SO(10) down to SU(5).
Adjoint 45 $A_{z}$ and $A_{u}$ are needed to break SU(5) further
down to the standard model
$SU(3)_{c} \times SU_{L} \times U(1)_{Y}$.

    The numerical predictions for the quark, lepton masses and quark
mixings are presented in  table 1b
with the input parameters and their values given in table 1a.
RG effects have been considered following the standard
scheme\cite{THIRD,OPERATOR} by
integrating the full two-loop RG equations from the GUT
scale down to the weak scale using
 $M_{SUSY} \simeq  M_{WEAK} \simeq M_{t} \simeq 180$GeV. From
the weak scale down to the lower energy scale,
three loops in QCD and two loops in QED are taken into consideration.
SUSY threshold effects are not considered in detail here since
the spectrum of sparticles is not yet determined.
The bottom quark mass may receive corrections as large
as $30 \%$ \cite{THRESHOLD} due to large $\tan \beta$.
However, it could be reduced by taking a suitable spectrum of superparticles.
Therefore, one should not expect to have
precise predictions untill the spectrum of the sparticles is
well determined. The strong coupling constant $\alpha_{s}(M_{Z})$ is
taken to be a free parameter with values given by the present experimental
bounds $\alpha_{s}(M_{z}) = 0.117 \pm 0.005$ \cite{PDG} in the following.

  {\bf Table 1a.}  Parameters
and their values as a function of the strong coupling $\alpha_{s}(M_{Z})$
determined by $m_{b}$, $m_{\tau}$, $m_{\mu}$, $m_{e}$ and $|V_{us}|= \lambda$.
\\

\begin{tabular}{|c|c|c|c|c|}  \hline
$\alpha_{s}(M_{Z})$   &  $\phi$ &
$\epsilon_{G} \equiv v_{5}/v_{10}$ & $\epsilon_{P}\equiv v_{5}/\bar{M}_{P}$
&  $\tan\beta $   \\   \hline
0.110  &  $73.4^{\circ}$  &  $2.66 \times
10^{-1}$  &   $ 0.89\times 10^{-2}$  & 51  \\
0.115    &  $77.5^{\circ}$  &  $2.51 \times
10^{-1}$  &   $ 0.83\times 10^{-2}$  & 55  \\
0.120    &  $81.5^{\circ}$  &  $2.34 \times
10^{-1}$  &   $ 0.77\times 10^{-2}$  & 58  \\   \hline
\end{tabular}
\\

{\bf  Table 1b.}  Observables and their predicted values with the values
of the parameters given in the table 1a.
\\

\begin{tabular}{|c|c|c|ccc|}   \hline
 Input   &     &  Output with $\alpha_{s}(M_{Z})$  &  0.110  &   0.115
&  0.120   \\ \hline
$m_{b}(m_{b})$\ [GeV]  &  4.35   &
$m_{t}$\ [GeV]  &  165   &  176  &  185  \\
$m_{\tau}$\ [GeV]  &  1.78   & $m_{c}(m_{c})$\ [GeV] & 1.14 & 1.30 & 1.37 \\
$m_{\mu}$\ [MeV]  &  105.6   &  $m_{s}$(1GeV)\ [MeV]
&  152   &  172 &  197  \\
$m_{e}$\ [MeV]  &  0.51 & $m_{d}$(1GeV)\ [MeV]
&  6.5   &  7.2  &  8.0  \\
$|V_{us}| \simeq \lambda$  & 0.22  &  $m_{u}$(1GeV)\ [MeV]
&  3.3  &  4.3  &  6.1  \\
 &   &  $|V_{cb}|\simeq A\lambda^{2}$
&  0.045   &  0.045  &  0.043  \\
 &  &  $|\frac{V_{ub}}{V_{cb}}|\simeq \lambda
\sqrt{\rho^{2} + \eta^{2}}$  & 0.053  & 0.056  & 0.063    \\
  &    &  $|\frac{V_{td}}{V_{cb}}|\simeq \lambda
\sqrt{(1-\rho)^{2} + \eta^{2}}$  & 0.201  &  0.199  & 0.198  \\  \hline
\end{tabular}
\\

 From  table 1a,
one sees that the model has large
$\tan\beta$ solution with
$\tan\beta \equiv v_{2}/v_{1} \sim m_{t}/m_{b}$.
CP violation is near maximum with a phase $\phi \sim 80^{\circ}$.
The vacuum structure between the GUT scale and Planck scale  has a hierarchic
structure $\epsilon_{G}\equiv v_{5}/v_{10} \sim \lambda = 0.22$ and
$\epsilon_{P}\equiv v_{5}/\bar{M}_{P} \sim \lambda^{3}$. Assuming
$(\bar{M}_{P}/M_{P})^{2} \simeq \alpha_{G} \simeq 1/24\sim \lambda^{2}$
(here $\alpha_{G}$ is the unified gauge coupling, $M_{P}$ is the
Planck mass),  we have
\begin{eqnarray}
& & \bar{M}_{P}=2.5\times 10^{18} GeV, \nonumber \\
& & v_{10} \simeq (0.86 \pm 0.16) \times 10^{17} GeV, \\
& & v_{5}\equiv M_{G} \simeq (2.2 \pm 0.2) \times 10^{16} GeV \nonumber
\end{eqnarray}
where the resulting value for the GUT scale agree well with the one
obtained from the gauge coupling unification.
$\bar{M}_{P}$ is also very close to the
reduced Planck scale $\hat{M}_{P} = M_{P}/\sqrt{8\pi} = 2.4 \times
10^{18}$ GeV and may be regarded as the scale for gravity unification.

It is seen  from table 1b that the predictions on
fermion masses and Cabbibo-Kobayashi-Maskawa (CKM)  mixing angles
fall in the range allowed by the experimental data\cite{PDG,TOP,MASS}:
\begin{equation}
 \begin{array}{lll}
m_{\tau}= 1777 MeV & m_{\mu}=105.6 MeV & m_{e} =0.51 MeV \\
m_{b}(m_{b}) = (4.15-4.35)GeV & m_{s}(1GeV) = (105-230)MeV & m_{d}(1GeV)
=(5.5-11.5)MeV \\
m_{t}(m_{t}) = (157-191)GeV & m_{c}(m_{c})= (1.22-1.32) GeV & m_{u}(1GeV) =
(3.1-6.4)MeV  \\
\end{array}
\end{equation}
and
\begin{equation}
V = \left(  \begin{array}{ccc}
V_{ud} & V_{us} & V_{ub} \\
V_{cd} & V_{cs} & V_{cb} \\
V_{td} & V_{ts} &  V_{tb} \\
\end{array} \right)
= \left(  \begin{array}{ccc}
0.9747-0.9759 & 0.218-0.224 & 0.002-0.005 \\
0.218-0.224  & 0.9738-0.9752 & 0.032-0.048 \\
0.004-0.015 & 0.03-0.048 & 0.9988-0.9995 \\
\end{array} \right)
\end{equation}
The model also gives a consistent prediction
for the $B^{0}$-$\bar{B}^{0}$ mixing and CP violation in kaon decays
(A detailed analysis will be presented elsewhere).

  It is of interest to expand the above fermion Yukawa
coupling matrices $\Gamma_{f}^{G}$ in terms of the
parameter $\lambda=0.22$ (the Cabbibo angle),
which was  found in \cite{WOLFENSTEIN} to be
very useful for expanding the CKM mixing matrix.
With the input values given in the table 1a, we find
\begin{eqnarray}
& & \Gamma_{u}^{G} \simeq \frac{2}{3}\lambda_{H} \left( \begin{array}{ccc}
0  & 0.97\lambda^{6}  & 0  \\
0.97\lambda^{6}  & 0  & -0.89\lambda^{2} \\
0  &  -0.89\lambda^{2}  &  1
\end{array}  \right); \  \Gamma_{d}^{G} \simeq \frac{2}{3}\lambda_{H}
\left( \begin{array}{ccc}
0  & -1.27\lambda^{4}  & 0  \\
-1.27\lambda^{4}  & 1.39 \lambda^{3}\  e^{i0.86\pi/2}  & -0.97\lambda^{3} \\
0  &  -0.97\lambda^{3}  &  1
\end{array}  \right) \\
& & \Gamma_{e}^{G} \simeq \frac{2}{3}\lambda_{H} \left( \begin{array}{ccc}
0  & -1.27\lambda^{4}  & 0  \\
-1.27\lambda^{4}  & 0.92 \lambda^{2}\  e^{i0.86\pi/2}  & 1.16\lambda^{2} \\
0  &  1.16\lambda^{2}  &  1
\end{array}  \right); \  \Gamma_{\nu}^{G} \simeq \frac{2}{3}\lambda_{H}
\left( \begin{array}{ccc}
0  & 0.86\lambda^{5}  & 0  \\
0.86\lambda^{5}  & 0.85 \lambda^{7}\  e^{i0.86\pi/2}  & -1.14\lambda^{6} \\
0  &  -1.14\lambda^{6}  &  1
\end{array}  \right) \nonumber
\end{eqnarray}
for $\alpha_{s}(M_{Z}) = 0.115$.

\section{Neutrino Masses and Mixings}

   To find the neutrino masses and mixings will be crucial tests of the model.
Many unification theories predict a see-saw type mass\cite{SSMASS}
$m_{\nu_{i}} \sim m_{u_{i}}^{2}/M_{N}$ with $u_{i} =u, c, t$ being
up-type quarks. For $M_{N} \simeq (10^{-3}\sim 10^{-4}) M_{GUT}
\simeq 10^{12}-10^{13}$ GeV, one has
\begin{equation}
m_{\nu_{e}} < 10^{-7} eV, \qquad m_{\nu_{\mu}} \sim 10^{-3} eV,
\qquad m_{\nu_{\tau}} \sim (3-21) eV
\end{equation}
in this case solar neutrino anomalous could be explained by
$\nu_{e} \rightarrow \nu_{\mu}$ oscillation, and the mass of
$\nu_{\tau}$ is in the range
relevant to hot dark matter.  However,  LSND events and atmospheric
neutrino deficit can not be explained in this scenario.

 By choosing Majorana type Yukawa coupling matrix differently, one can
construct many models of neutrino mass matrix. We shall present one here,
which is found to be of interest with the following texture:
\begin{equation}
M_{N}^{G} = \lambda_{H} v_{10} \frac{\epsilon_{P}^{4}}{\epsilon_{G}^{2}}
\left( \begin{array}{ccc}
0  &  0 &   \frac{1}{2}z_{N}   \\
0  &  y_{N} & 0 \\
\frac{1}{2}z_{N}   & 0 &  w_{N}
\end{array} \right)
\end{equation}
 The corresponding effective operators are given by
\begin{eqnarray}
W_{33}^{N} & = & \lambda_{H} \frac{v_{10}}{2} \frac{\epsilon_{P}^{4}}{
\epsilon_{G}^{2}} 16_{3} (\frac{A_{u}}{v_{5}}) (\frac{\bar{\Phi}}{
v_{10}/\sqrt{2}})(\frac{\bar{\Phi}}{
v_{10}/\sqrt{2}}) (\frac{A_{B-L}}{v_{5}}) 16_{3}  \nonumber \\
W_{13}^{N} & = & \lambda_{H} \frac{v_{10}}{2} \frac{\epsilon_{P}^{4}}{
\epsilon_{G}^{2}} 16_{1} (\frac{A_{u}}{v_{5}}) (\frac{\bar{\Phi}}{
v_{10}/\sqrt{2}})(\frac{\bar{\Phi}}{
v_{10}/\sqrt{2}}) (\frac{A_{3R}}{v_{5}}) 16_{3}  \\
W_{22}^{N} & = & \lambda_{H} \frac{v_{10}}{2} \frac{\epsilon_{P}^{4}}{
\epsilon_{G}^{2}} 16_{2} (\frac{A_{u}}{v_{5}}) (\frac{\bar{\Phi}}{
v_{10}/\sqrt{2}})(\frac{\bar{\Phi}}{
v_{10}/\sqrt{2}}) (\frac{A_{u}}{v_{5}}) 16_{2} \nonumber
\end{eqnarray}
where $w_{N}$, $y_{N}$ and $z_{N}$ are Clebsch factors with
$w_{N} = 4/3$, $y_{N} = 16/9$,  $z_{N} = 2/3$.  They are determined by
additional 45s $A_{B-L}$ and $A_{3R}$ with
$<A_{B-L}> =v_{5}\  diag. (\frac{2}{3},\ \frac{2}{3},\ \frac{2}{3},\
0,\ 0)\otimes \tau_{2}$ and
$<A_{3R}> =v_{5}\  diag. (0,\ 0,\ 0,\
\frac{1}{2},\ \frac{1}{2})\otimes \tau_{2}$.
The 45 $A_{B-L}$ is also
necessary for doublet-triplet mass splitting\cite{DW} in the Higgs $10_{1}$.

 The light neutrino mass matrix is given via see-saw mechanism as
follows
\begin{eqnarray}
M_{\nu} = \Gamma_{\nu}^{G} (M_{N}^{G})^{-1}
(\Gamma_{\nu}^{G})^{\dagger} v_{2}^{2}
& = & M_{0} \left( \begin{array}{ccc}
\frac{3}{4}\frac{z_{N}}{|z_{\nu}|}\frac{1}{y_{N}}  &
\frac{3}{2}\frac{z_{N}}{z_{\nu}^{2}} \frac{y_{\nu}}{y_{N}}
\frac{\epsilon_{G}^{2}}{\epsilon_{P}^{2}} e^{-i\phi} &
-\frac{\sqrt{3}}{4} \frac{z_{N}}{|z_{\nu}|} \frac{x_{\nu}}{y_{N}}
\frac{\epsilon_{G}^{2}}{\epsilon_{P}}   \\
\frac{3}{2}\frac{z_{N}}{z_{\nu}^{2}} \frac{y_{\nu}}{y_{N}}
\frac{\epsilon_{G}^{2}}{\epsilon_{P}^{2}} e^{i\phi}  &
-3 \frac{w_{N}}{|z_{\nu}|z_{N}} - \sqrt{3} \frac{x_{\nu}}{|z_{\nu}|}
\frac{\epsilon_{G}^{2}}{\epsilon_{P}} & 1 \\
-\frac{\sqrt{3}}{4} \frac{z_{N}}{|z_{\nu}|} \frac{x_{\nu}}{y_{N}}
\frac{\epsilon_{G}^{2}}{\epsilon_{P}}   & 1 &  \frac{3}{2}
\frac{z_{N}}{z_{\nu}^{2}} \frac{y_{\nu}}{y_{N}}
\frac{\epsilon_{G}^{2}}{\epsilon_{P}^{2}}
\end{array} \right) \nonumber \\
& = & 2.1\ \lambda_{H} \left( \begin{array}{ccc}
0.73 \lambda^{6}  &  0.73 \lambda^{8} e^{-i0.86\pi/2} &
-0.97 \lambda^{7}  \\
0.73 \lambda^{8} e^{i0.86\pi/2}  &   -0.86\lambda^{4} & 1 \\
-0.97 \lambda^{7}  & 1 &  0.73 \lambda^{8}
\end{array} \right)
\end{eqnarray}
with $M_{0} = \frac{2}{3} \frac{\epsilon_{G}^{2}}{\epsilon_{P}^{4}}
\frac{1}{|z_{\nu}|z_{N}}\frac{1}{\eta_{\nu}} \frac{v_{2}}{v_{10}}v_{2}
\lambda_{H} $. Here $\eta_{\nu}$ is the RG evolution factor and estimated
to be $\eta_{\nu} \simeq 1.35$. Diagonalizing the above mass matrix,
we obtain masses of light Majorana neutrinos:
\begin{eqnarray}
\frac{m_{\nu_{e}}}{m_{\nu_{\mu}}} & = & \frac{3}{4}
\frac{z_{N}}{|z_{\nu}|} \frac{1}{y_{N}} = 0.83 \times 10^{-4}, \nonumber \\
\frac{m_{\nu_{\mu}}}{m_{\nu_{\tau}}} & = & 1 - 3
\frac{w_{N}}{|z_{\nu}|z_{N}} - \sqrt{3} \frac{x_{\nu}}{|z_{\nu}|}
\frac{\epsilon_{G}^{2}}{\epsilon_{P}} \simeq 0.998  \\
m_{\nu_{\tau}} & \simeq & M_{0} \simeq 2.1\  \lambda_{H}\  eV \nonumber
\end{eqnarray}
The three heavy Majorana neutrinos have masses
\begin{eqnarray}
\frac{m_{N_{1}}}{m_{N_{2}}} & = &
\frac{z_{N}^{2}}{4w_{N}y_{N}}  = 0.047, \qquad
\frac{m_{N_{2}}}{m_{N_{3}}} =
\frac{y_{N}}{w_{N}}  = 1.33  \nonumber \\
m_{N_{3}} & = & \frac{\epsilon_{P}^{4}}{\epsilon_{G}^{2}} w_{N} v_{10}
\lambda_{H} \simeq  0.64 \times 10^{10}\  \lambda_{H}\ GeV
\end{eqnarray}
The CKM-type lepton mixing matrix is predicted to be
\begin{equation}
V_{LEP} = V_{\nu} V_{e}^{\dagger}  = \left(  \begin{array}{ccc}
V_{\nu_{e}e} & V_{\nu_{e}\mu} & V_{\nu_{e}\tau} \\
V_{\nu_{\mu}e } & V_{\nu_{\mu}\mu} & V_{\nu_{\mu}\tau} \\
V_{\nu_{\tau}e} & V_{\nu_{\tau}\mu} &  V_{\nu_{\tau}\tau} \\
\end{array} \right) =
\left(  \begin{array}{ccc}
0.9976 & 0.068 & 0.000 \\
-0.051 & 0.748 & 0.665 \\
0.045 & -0.664 &  0.748 \\
\end{array} \right)
\end{equation}
CP-violating effects are found to be small in the lepton mixing matrix.
As a result we find the following:

1. a $\nu_{\mu}(\bar{\nu}_{\mu}) \rightarrow \nu_{e} (\bar{\nu_{e}})$
short wave-length oscillation with
\begin{equation}
\Delta m_{e\mu}^{2} = m_{\nu_{\mu}}^{2} - m_{\nu_{e}}^{2}
\simeq (4-6) eV^{2}, \qquad
\sin^{2}2\theta_{e\mu} \simeq 1.8 \times 10^{-2} \ ,
\end{equation}
which is consistent with the LSND experiment\cite{LSND}
\begin{equation}
\Delta m_{e\mu}^{2} = m_{\nu_{\mu}}^{2} - m_{\nu_{e}}^{2}
\simeq (4-6) eV^{2}\ , \qquad
\sin^{2}2\theta_{e\mu} \simeq 1.8 \times 10^{-2} \sim 3 \times 10^{-3}\ ;
\end{equation}

2. a $\nu_{\mu} (\bar{\nu}_{\mu}) \rightarrow \nu_{\tau} (\bar{\nu}_{\tau})$
long-wave length oscillation with
\begin{equation}
\Delta m_{\mu\tau}^{2} = m_{\nu_{\tau}}^{2} - m_{\nu_{\mu}}^{2} \simeq
(1.6-2.4)\times 10^{-2} eV^{2}\ , \qquad
\sin^{2}2\theta_{\mu\tau} \simeq 0.987 \ ,
\end{equation}
which could explain the atmospheric neutrino
deficit\cite{ATMO}:
\begin{equation}
\Delta m_{\mu\tau}^{2} = m_{\nu_{\tau}}^{2} - m_{\nu_{\mu}}^{2} \simeq
(0.5-2.4) \times 10^{-2} eV^{2}\ , \qquad
\sin^{2}2\theta_{\mu\tau} \simeq 0.6 - 1.0 \ ,
\end{equation}
with the best fit\cite{ATMO}
\begin{equation}
\Delta m_{\mu\tau}^{2} = m_{\nu_{\tau}}^{2} - m_{\nu_{\mu}}^{2} \simeq
1.6\times  10^{-2} eV^{2}\ , \qquad
\sin^{2}2\theta_{\mu\tau} \simeq 1.0 \ ;
\end{equation}
However, $(\nu_{\mu} - \nu_{\tau})$ oscillation will be beyond the reach
of CHORUS/NOMAD and E803.

3. Two massive neutrinos $\nu_{\mu}$ and $\nu_{\tau}$ with
\begin{equation}
m_{\nu_{\mu}} \simeq m_{\nu_{\tau}}  \simeq (2.0-2.4) eV \ ,
\end{equation}
which fall in the range required by
possible hot dark matter\cite{DARK}.

  In this case, solar neutrino deficit has to be explained by oscillation
between $\nu_{e}$ and  a sterile neutrino\cite{STERILE} $\nu_{s}$. Since
strong bounds on the number of neutrino species both from the invisible
$Z^{0}$-width and from primordial nucleosynthesis \cite{NS,NS1} require the
additional neutrino to be sterile (singlet of $SU(2)\times U(1)$, or
singlet of SO(10) in the GUT SO(10) model).
Masses and mixings of the triplet sterile neutrinos can be chosen
by introducing an additional  singlet scalar with VEV $v_{s}\simeq 450$ GeV.
We find
\begin{eqnarray}
& & m_{\nu_{s}} = \lambda_{H} v_{s}^{2}/v_{10} \simeq 2.4 \times 10^{-3} eV
\nonumber \\
& & \sin\theta_{es} \simeq \frac{m_{\nu_{L}\nu_{s}}}{m_{\nu_{s}}}
= \frac{v_{2}}{2v_{s}} \frac{\epsilon_{P}}{\epsilon_{G}^{2}} \simeq 3.8
\times 10^{-2}
\end{eqnarray}
with the mixing angle  consistent with the requirement necessary for
primordial nucleosynthesis \cite{PNS}
given by \cite{NS}.  The resulting parameters
\begin{equation}
\Delta m_{es}^{2} = m_{\nu_{s}}^{2} - m_{\nu_{e}}^{2} \simeq 5.8 \times
10^{-6} eV^{2}, \qquad  \sin^{2}2 \theta_{es} \simeq 5.8 \times 10^{-3}
\end{equation}
are consistent with the values \cite{STERILE} obtained from
fitting the experimental data:
\begin{equation}
\Delta m_{es}^{2} = m_{\nu_{s}}^{2} - m_{\nu_{e}}^{2} \simeq (4-9) \times
10^{-6} eV^{2}, \qquad  \sin^{2}2 \theta_{es} \simeq (1.6-14) \times 10^{-3}
\end{equation}

  This scenario can be tested by the next generation solar neutrino
experiments in Sudhuray Neutrino Observatory (SNO) and
Super-kamiokanda (Super-K), both planning to start
operation in 1996. From measuring  neutral current events, one could
identify $\nu_{e} \rightarrow \nu_{s}$ or
$\nu_{e} \rightarrow \nu_{\mu} (\nu_{\tau})$ since the sterile
neutrinos have no weak gauge interactions. From measuring seasonal
variation, one can further distinguish the small-angle MSW \cite{MSW}
oscillation from vacuum mixing oscillation.

\section{Superpotential for Fermion Yukawa Interactions}

  Non-Abelian discrete family symmetry $\Delta (48)$ is important in the
present model for constructing interesting texture structures of the Yukawa
coupling matrices. It initiates from basic considerations that all three
families are treated on the same footing at the GUT scale, namely the
three families should belong to an irreduciable triplet representation of
a family symmetry group. Based on the well-known fact that masses of the
three families have a hirarchic structure,  the family symmetry group must be
a group with at least rank three if the group is a continuous one. However,
within the known simple continuous groups, it is difficult to find a rank three
group which has irreduciable triplet representations. This limotation of the
continuous groups is thus avoided by their finite and disconnected subgroups.
A simple example is the finite and disconnected group $\Delta (48)$, a subgroup
of $SU(3)$.

   The generators of the $\Delta (3n^{2})$ group consist of the matrices

\begin{equation}
E(0,0) = \left( \begin{array}{ccc}
0 & 1 & 0 \\
0 & 0 & 1 \\
1 & 0 & 0  \\
\end{array}   \right)
\end{equation}
and
\begin{equation}
A_{n}(p,q) = \left( \begin{array}{ccc}
e^{i\frac{2\pi}{n}p} & 0 & 0 \\
0 & e^{i\frac{2\pi}{n}q} & 0 \\
0 & 0 & e^{-i\frac{2\pi}{n}(p+q)}  \\
\end{array}   \right)
\end{equation}

 It is clear that there are $n^2$ different elements $A_{n}(p,q)$ since if $p$
is fixed, $q$ can take on $n$ different values. There are three different
elements types,

\[   A_{n}(p,q), \qquad E_{n}(p,q)=A_{n}(p,q)E(0,0), \qquad
C_{n}(p,q)=A_{n}(p,q)E^{2}(0,0)  \]
in the $\Delta (3n^{2})$ group, therefore the order of the
$\Delta (3n^{2})$ group is $3n^{2}$.  The irreducible
representations of the $\Delta (3n^{2})$
groups consist of i) $(n^2 -1)/3$ triplets and three
singlets when $n/3$ is not an interger and ii) $(n^2 -3)/3$ triplets and nine
singlets when $n/3$ is an interger.

   The characters of the triplet representations can be expressed
\cite{SUBGROUP}

\begin{eqnarray}
\Delta_{T}^{m_{1}m_{2}} (A_{n}(p,q)) & = & e^{i\frac{2\pi}{n}[m_{1}p + m_{2}q]}
+ e^{i\frac{2\pi}{n}[m_{1}q - m_{2}(p+q)]} +
e^{i\frac{2\pi}{n}[-m_{1}(p+q) + m_{2}p]}  \\
\Delta_{T}^{m_{1}m_{2}} (E_{n}(p,q)) & = & \Delta_{T}^{m_{1}m_{2}} (C_{n}(p,q))
=0  \nonumber
\end{eqnarray}
with $m_{1}$, $m_{2} = 0,1, \cdots ,  n-1$. Note that $(-m_{1}+m_{2} , -m_{1})$
and $(-m_{2} , m_1 - m_2 )$ are equivalent to $(m_1 , m_{2})$.

  One will see that $\Delta (48)$ (i.e., n=4 )
is the smallest of the dihedral
group $\Delta (3n^{2})$  with sufficient triplets for
constructing interesting texture structures of the Yukawa coupling matrices.

   The irreducible triplet representations of $\Delta (48)$ consist of
two complex triplets $T_{1}(\bar{T}_{1})$ and $T_{3}(\bar{T}_{3})$ and one real
triplet $T_{2} = \bar{T}_{2}$ as well as three singlet representations.
Their irreducible triplet representations can be expressed
in terms of the matrix representation
\begin{eqnarray}
T_{1}^{(1)} & = & diag. (i, 1, -i), \qquad T_{1}^{(2)}  = diag. (1, -i, i),
\qquad T_{1}^{(3)}  = diag. (-i, i, 1); \nonumber  \\
T_{2}^{(1)} & = & diag. (-1, 1, -1), \qquad T_{2}^{(2)}  = diag. (1, -1, -1),
\qquad T_{2}^{(3)}  = diag. (-1, -1, 1);  \\
T_{3}^{(1)} & = & diag. (i, -1, i), \qquad T_{3}^{(2)}  = diag. (-1, i, i),
\qquad T_{3}^{(3)}  = diag. (i, i, -1) \nonumber
\end{eqnarray}
The matrix representations of $\bar{T}_{1}^{(i)}$ and $\bar{T}_{3}^{(i)}$ are
the hermician conjugates of $T_{1}^{(i)}$ and $T_{3}^{(i)}$.
{}From this representation, we can explicitly construct the invariant tensors.

{\bf Table 2},  Decomposition of the product of two triplets,
$T_{i}\otimes T_{j}$ and $T_{i}\otimes \bar{T}_{j}$ in $\Delta_{48}(SU(3))$.
Triplets $T_{i}$ and $\bar{T}_{i}$ are simply denoted by $i$ and $\bar{i}$
respectively. For example $T_{1}\otimes \bar{T}_{1} = A \oplus T_{3} \oplus
\bar{T}_{3} \equiv A3\bar{3}$, here $A$ represents a singlet.
\\

\begin{tabular}{|c|ccccc|}  \hline
$\Delta (48)$ & 1   & $\bar{1}$  &  2  &  3  & $\bar{3}$  \\ \hline
  1        & $\bar{1} \bar{1}$2  & A3$\bar{3}$  & $\bar{1} 3 \bar{3}$ &
 123  &
12$\bar{3}$  \\
2   &  $\bar{1}3\bar{3}$  & 13$\bar{3}$   &  A23  & 1$\bar{1}\bar{3}$  &
1$\bar{1}$3 \\
3  &   123  & $\bar{1}$23  & 1$\bar{1}\bar{3}$  & 2$\bar{3}\bar{3}$  &
A1$\bar{1}$ \\    \hline
\end{tabular}
\\

  All three families with $3\times 16 = 48$
chiral fermions are unified into a triplet 16-dimensional spinor
representation of $SO(10)\times \Delta (48)$.
Without losing generality, one can assign the three
chiral families into the triplet representation $T_{1}$, which may be simply
denoted as $\hat{16} = 16_{i} T_{1}^{(i)}$.
All the fermions are assumed to obtain their masses through a single
$10_{1}$ of SO(10) into which the needed two Higgs doublets are unified.
The model could allow a triplet sterile neutrino with small mixings with
the ordinary neutrinos. A singlet scalar near the electroweak scale
is necessary to generate small masses for the sterile neutrinos.

 Superpotentials which lead to the above
texture structures (eqs. (1), (2) and (9)) with zeros and
effective operators (eqs. (3) and (10)) are found to be
\begin{eqnarray}
W_{Y} & = & \sum_{a=0}^{3} \psi_{a1} 10_{1} \psi_{a2} +
\bar{\psi}_{22} \chi \psi_{13} + \bar{\psi}_{21} \chi_{2} \psi_{13} +
\bar{\psi}_{32} \chi \psi_{23} + \bar{\psi}_{31} \chi_{3} \psi_{23} \nonumber
\\
& & + \bar{\psi}_{02} \chi \psi_{33} + \bar{\psi}_{01} \chi_{0} \psi_{33} +
\bar{\psi}_{33} A_{X} \psi_{3} + \bar{\psi}_{3} A_{X} \psi_{2} +
\bar{\psi}_{2} A_{X} \psi_{1}  \nonumber \\
& & + (\bar{\psi}_{11} \chi_{1}  + \bar{\psi}_{12} \chi  +
\bar{\psi}_{13} A_{z}  + \bar{\psi}_{23} A_{u}   +
\bar{\psi}_{1} Y ) \hat{16}  \\
& & + \sum_{a=0}^{3} \sum_{j=1}^{3} S_{G} \bar{\psi}_{aj} \psi_{aj}
+ \sum_{i=1}^{2} (\bar{\psi}_{i3} A_{X} \psi_{i3} + S_{I} \bar{\psi}_{i}
\psi_{i} ) + S_{I} \bar{\psi}_{33} \psi_{33} + S_{P} \bar{\psi}_{3}\psi_{3}
\nonumber
\end{eqnarray}
for the fermion Yukawa coupling matrices, and
\begin{eqnarray}
W_{R} & = & \sum_{i=1}^{3} (\psi'_{i1} 10_{3} \psi'_{i2} +
\bar{\psi}'_{i1} \chi'_{i} \psi + \bar{\psi}'_{i2} \chi' \psi'_{i3} +
 \bar{\psi}'_{i1} A_{i} \psi' ) +
(\bar{\psi}' X  + \bar{\psi} A_{u} ) \hat{16} + \bar{\Phi} 10_{3}
\bar{\Phi}  \nonumber  \\
& &  + \sum_{i=1}^{3} \sum_{j=1}^{2} S_{G} \bar{\psi}'_{ij} \psi'_{ij}
+ S_{P} (\sum_{i=1}^{3} \bar{\psi}'_{i3} \psi'_{i3} + \bar{\psi} \psi +
\bar{\psi}' \psi'  ) \nonumber
\end{eqnarray}
for the right-handed Majorana neutrinos, and
\begin{equation}
W_{S} =  \bar{\psi}'_{1}10_{1} \psi'_{2} + \bar{\psi}'_{1} \Phi \nu_{s} +
\bar{\psi}'_{2} \phi_{s} \hat{16} +
(\bar{\nu}_{s} \phi_{s} N_{s} + h.c. ) + S_{I} \bar{N}_{s} N_{s}
\end{equation}
for the sterile neutrino masses and their mixings with the ordinary neutrinos.

   In the above superpotentials,  each term is ensured by the $U(1)$ symmetry.
An appropriate assignment of U(1) charges for the various fields is implied.
All $\psi$ fields are triplet
16-D spinor heavy fermions.
Where the fields $\psi_{i3} \{\bar{\psi}_{i3}\}$,
$\psi_{i3}' \{\bar{\psi}'_{i3}\}$,
$\psi_{i} \{\bar{\psi}_{i}\}$ ($i = 1,2,3$), $\psi'_{1} \{\bar{\psi}'_{1}\}$,
$\psi'_{2} \{\bar{\psi}'_{2}\}$,
$\psi \{\bar{\psi}\}$ and $\psi' \{\bar{\psi}'\}$
belong to $(16, T_{1})\{(\bar{16}, \bar{T}_{1})\}$ representations
of $SO(10) \times \Delta_{48}(SU(3)) $; $\psi_{11}\{\bar{\psi}_{11}\}$ and
$\psi_{12}\{\bar{\psi}_{12}\}$ belong to $(16, T_{2})\{(\bar{16}, T_{2})\}$;
$\psi_{i1}\{\bar{\psi}_{i1}\}$  and $\bar{\psi}_{i2}\{\psi_{i2}\}$ ($i=2,3,0$)
belong to $(16,T_{3}) \{(\bar{16}, \bar{T}_{3})\}$;
$\psi'_{i1}\{\bar{\psi}'_{i1}\}$  and $\bar{\psi}'_{i2}\{\psi'_{i2}\}$
($i=1,2$) belong to $(16, \bar{T}_{3}) \{(\bar{16}, T_{3})\}$;
$\psi'_{31}\{\bar{\psi}'_{31}\}$ and
$\psi'_{32}\{\bar{\psi}'_{32}\}$ belong to
$(16, T_{2})\{(\bar{16}, T_{2})\}$;
$X$, $Y$, $S_{I}$, $S_{P}$ and $\phi_{s}$ are singlets of $SO(10) \times
\Delta_{48}(SU(3))$.  $\nu_{s}$ and $N_{s}$ are SO(10) singlet
and $\Delta (48)$ triplet fermions.  $10_{3}$ is an additional
SO(10) 10-representation heavy scalar.
All SO(10) singlet $\chi$ fields are triplets of
$\Delta (48)$. Where $(\chi_{1}, \chi_{2}, \chi_{3}, \chi_{0}, \chi)$
belong to triplet representations $(\bar{T}_{3}, T_{3}, \bar{T}_{1}, T_{2},
\bar{T}_{3})$ respectively; $(\chi'_{1}, \chi'_{2}, \chi'_{3}, \chi')$
belong to triplet representations $(\bar{T}_{1},  T_{2}, T_{3},  T_{3})$
respectively. With the above assignment for various fields,  one can check
that once the triplet field  $\chi$ develops
VEV only along the third direction, i.e.,
$<\chi^{(3)}> \neq 0$, and $\chi'$ develops
VEV only along the second direction, i.e.,
$<\chi'^{(2)}> \neq 0$,
the resulting fermion Yukawa coupling matrices at the GUT scale will be
automatically forced, due to the special features of
$\Delta (48)$, into an interesting texture structure with four
non-zero textures `33', `32', `22' and `12' which are characterized
by $\chi_{1}$, $\chi_{2}$, $\chi_{3}$, and $\chi_{0}$ respectively, and
the resulting right-handed Majorana neutrino mass matrix is forced into
three non-zero textures `33', `13' and `22' which are characterized
by $\chi'_{1}$, $\chi'_{2}$, and $\chi'_{3}$ respectively.
It is seen that five triplets are needed. Where one triplet is necessary for
unity of the three family fermions, and four triplets are required for
obtaining the needed minimal non-zero textures.

  The symmetry breaking scenario and the structure of the physical vacuum are
considered as follows
\begin{eqnarray}
& & SO(10)\times \Delta (48) \times U(1) \stackrel{\bar{M}_{P}}{\rightarrow}
SO(10)\times \Delta (48) \stackrel{v_{10}}{\rightarrow} SU(5)\times \Delta (48)
\nonumber \\
& & \stackrel{v_{5}}{\rightarrow} SU(3)_{c}\times SU(2)_{L} \times U(1)_{Y}
\stackrel{v_{1}, v_{2}}{\rightarrow} SU(3)_{c}\times U(1)_{em}
\end{eqnarray}
and: $<S_{P}> = \bar{M}_{P}$,
$<X> = v_{10} = <S_{I}> $,  $<\Phi^{(16)}> = <\bar{\Phi}^{(16)}>
= v_{10}/\sqrt{2}$, $<Y> = v_{5} = <S_{G}>$ ,
$<\chi^{(3)}> = <\chi^{(i)}_{a}> = <\chi'^{(2)}> = <\chi'^{(i)}_{j}> = v_{5}$
 with $(i=1,2,3; a=0,1,2,3; j=1,2,3)$,
$<\chi^{(1)}> = <\chi^{(2)}> =<\chi'^{(1)}> =
<\chi'^{(3)}> =0 $,  $<\phi_{s}> = v_{s} \simeq 450$ GeV,
$<H_{2}> = v_{2} = v \sin\beta $ with
$v = \sqrt{v_{1}^{2} + v_{2}^{2}}= 246 $ GeV.

\section{Conclusions}

 It is amazing that nature has allowed us to make predictions in terms
of a single Yukawa coupling constant and a set of VEVs determined by
the structure of the vacuum
and to understand the low energy physics
from the Planck scale physics. The present model has provided
a consistent picture on the 28 parameters in SM model with massive neutrinos.
The neutrino sector is of special interest to further study.
Though the recent LSND experiment, atmospheric neutrino deficit,
and hot dark matter could be simultaneously explained
in the present model, however, solar neutrino puzzle can be
understood  by introducing an SO(10) singlet sterile neutrino.
It is expected that more precise measurements from various low energy
experiments in the near future could provide
crucial tests on the present model.
\\

{\bf ACKNOWLEDGEMENT}

 YLW would like to thank Inst. for Theor. Physics, Chinese Academy of Sciences,
for its hospitality and partial support during his visit.

\end{document}